\documentclass[12pt,aps,prb,groupedaddress,showpacs,showkeys,preprint,tightenlines]{revtex4}
\usepackage{amssymb,amsmath}
\usepackage{graphicx}
\usepackage{dcolumn}
\usepackage{bm}

\begin{document}

\title{Band structures of P-, D-, and G-surfaces}

\author{Nobuhisa Fujita}
\email[]{nobuhisa@struc.su.se}
\affiliation{Department of Structural Chemistry, Arrhenius Laboratory,
Stockholm University, 10691 Stockholm, Sweden}

\author{Osamu Terasaki}
\affiliation{Department of Structural Chemistry, Arrhenius Laboratory,
Stockholm University, 10691 Stockholm, Sweden}

\date{\today}

\begin{abstract}
We present a theoretical study on the band structures of the electron
constrained to move along triply-periodic minimal surfaces. Three well
known surfaces connected via Bonnet transformations, namely P-, D-, and 
G-surfaces, are considered. The six-dimensional algebra of the Bonnet 
transformations [C. Oguey and J.-F. Sadoc, J. Phys. I France {\bf 3}, 
839 (1993)] is used to prove that the eigenstates for these surfaces are
interrelated at a set of special points in the Brillouin zones. The global 
connectivity of the band structures is, however, different due to the 
topological differences of the surfaces. A numerical investigation of the 
band structures as well as a detailed analysis on their symmetry properties 
is presented. It is shown that the presence of nodal lines are closely
related to the symmetry properties. The present study will provide a basis 
for understanding further the connection between the topology and the band 
structures.
\end{abstract}



\maketitle

\section{Introduction}

For nearly two decades intense efforts have been devoted to the studies of 
periodic curved surfaces as topological objects to describe certain 
structured materials. Several conceptually different classes, such as 
zero-potential surfaces, nodal surfaces, and periodic minimal surfaces, 
often share similar properties concerning topology and 
symmetry.\cite{SN91,AHLL88} Such a surface can be constructed as a space 
partitioner which is invariant under space group operations, hence the 
space group symmetries are the key ingredients of them. In reality, on the 
other hand, such surfaces are realized as nanometer to micron scale 
interface structures in lyotropic liquid crystals or cell membranes. 
Furthermore, recent progress in the synthesis of nano-structured materials 
prompts us to consider the possibility of using these materials as the 
templates for electronic devices embracing interesting geometrical properties.

The present paper deals with a fundamental problem on the electronic 
structures of materials characterized by periodic curved surfaces. 
The theoretical basis of the quantum mechanics of electrons confined 
within thin curved films has been developed by several 
authors.\cite{JK71,C81,C82,IN91,DE95,CEK04} They showed that when the 
electron is strongly constrained onto a smoothly curved surface, the 
quantization of the motion perpendicular to the surface results in an 
effective potential energy depending on the local curvatures along the 
surface. Since this potential energy is attractive, the electron can be 
weakly bounded around a curved region of the surface (geometrically 
induced bound states). Theoretical interests also arise in regard to 
quantum chaos, since the constrained motion on a surface with non-zero 
intrinsic curvatures entails classical chaos.\cite{BV86} These studies, 
however, deal primarily with the local effects of curvatures. The roles 
played by the symmetry and topology of the surface on the determination 
of electronic structures have largely remained intact, apart from that 
Aoki et al. numerically studied the band structures of triply periodic 
minimal surfaces recently.\cite{AKTMK01,KA05} The aim of the present 
paper is to perform a more thorough investigation into the electronic 
structures of similar periodic curved surfaces.

In Sec.\ref{TPMS}, we provide a mathematical background for periodic minimal 
surfaces and introduce the three basic surfaces in the same Bonnet family, 
i.e., P-, D-, and G-surfaces, which will be studied further. 
In Sec.\ref{6Ddescription}, the translational subgroups of these three 
surfaces are described as orthogonal projections of an identical 
six-dimensional (6D) translational subgroup onto 3D subspaces. A numerical 
investigation of the band structures of the three surfaces is presented 
in Sec.\ref{bandstruc} following a brief summary of the basic formulation 
of the one-electron problem constrained onto a surface embedded in 3D space. 
In Sec.\ref{6Dband} the algebraic information given in 
Sec.\ref{6Ddescription} is used to prove certain interrelations between the 
band structures of the three surfaces. More precisely, we propose the idea 
of ``band intersections'' to show that the energy eigenvalues at a certain 
set of wave vectors {\em coincide} among the three surfaces. 
In Sec.\ref{symmetry} the symmetry properties of energy bands, i.e., the 
corresponding irreducible representations (IR's) of the space groups, are 
analyzed. Based on the IR's, several features of the band structures and 
eigenstates are discussed, including the band sticking phenomena and the 
occurrence of nodal lines. In particular, a general connection is 
established between the symmetry properties (or IR's) of the eigenstates 
and the types of nodal lines entailed by symmetry. The exhaustive lists of 
IR's with such nodal lines are shown. Further discussions are given in the 
last section.

\section{Periodic minimal surfaces}\label{TPMS}

Curved surfaces are naturally realized as interfaces, and their formation is 
dominated by surface energies (e.g. surface tension). Like soap films on 
closed wire frames, they often resemble minimal surfaces, that is, surfaces 
with vanishing mean-curvature. Minimal surfaces are privileged 
mathematically since they have an analytic representation called the 
Weierstrass-Enneper (WE) representation:\cite{S75,O86} a minimal surface is 
represented by the real part of a complex vector ${\bf z}$ in a 3D complex 
space ${\bf C}^3 \,(\simeq E_6)$ multiplied by a phase factor:
\begin{eqnarray}
{\bf r}(w) = {\rm Re}\left(e^{i\alpha}{\bf z}(w)\right),\label{WEformula}
\end{eqnarray}
\begin{eqnarray}
{\bf z}(w) = \ell\left(
\int^{w}\!\! F(\zeta)(1-\zeta^2){\rm d}\zeta,\,
\int^{w}\!\! i F(\zeta)(1+\zeta^2){\rm d}\zeta,\,
\int^{w}\!\! 2 F(\zeta) \zeta {\rm d} \zeta\right),
\label{WEintegral}
\end{eqnarray}
where $\ell$ is an overall scale factor, $\alpha$ the Bonnet angle, 
$w=q^1+i q^2$, and $F(\zeta)$ a complex analytic (or holomorphic) function 
associated with the Bonnet family of minimal surfaces.\cite{BonnetFamily}
Hence the surface is represented as a multiple-valued function of the 
complex variable $w$, where different integration paths give distinct points 
on the surface. The two real variables, $q^1$ and $q^2$, serve as the 
isothermal parameterization of the surface.

Minimal surfaces with periodicity in three independent directions are called 
triply periodic minimal surfaces (TPMS's). The importance of TPMS's as 
topological objects describing interface materials like self-assembled 
lyotropic liquid crystals has been recognized for nearly two 
decades.\cite{AHLL88,CS87,SC89} Compared to other classes of periodic 
surfaces like zero-potential surfaces,\cite{SN91} TPMS's are especially 
advantageous for physical applications because Eq.(\ref{WEformula}) 
provides a parameterization of the surface for incorporating it into 
written equations. In the following, we focus on the classical examples of 
Schwarz's P- and D-surfaces\cite{Schwarz} and Sch$\ddot{\rm o}$n's 
G-surface (gyroid)\cite{Schoen} which belong to the cubic system. 
The corresponding WE function, $F(\zeta)$, is given by
\begin{eqnarray}
F(\zeta)&=&\frac{1}{\sqrt{1+14\zeta^4+\zeta^8}},
\end{eqnarray}
with eight branch points of order 2 at 
$\zeta=(\sqrt{3}\pm 1)/\sqrt{2}e^{2\pi i n/4}$ ($n=0,1,2,3$) on its Riemann 
surface.\cite{RiemannSurface} For each of the three surfaces, the Bonnet 
angle $\alpha$, has to take a proper value as will be given below. 
Multiple surfaces in the same Bonnet family are intrinsically identical, 
or isometric, apart from the global topology. The surfaces are obtained from 
one another via Bonnet transformation, namely the change of $\alpha$.

We may write ${\bf z}(w) = {\bf u}(w) + i {\bf v}(w)$ with ${\bf u}(w)$ 
and ${\bf v}(w)$ being two real vectors, hence 
${\bf r}(w) = {\bf u}(w) \cos{\alpha} - {\bf v}(w)\sin{\alpha}$. 
Alternatively, one may introduce a 6D vector, 
$\Xi(w) = ({\bf u}(w), \, {\bf v}(w))$, and the minimal surface ${\bf r}(w)$ 
is given as the orthogonal projection of $\Xi(w)$ onto a 3D subspace 
specified by the hyper-rotation angle $\alpha$. P- and D-surfaces are given
by the Bonnet angles, $\alpha_P=0$ and $\alpha_D=90^{\circ}$, hence they 
are given by ${\bf u}(w)$ and $-{\bf v}(w)$, respectively. Accordingly the 
6D space $E_6$ embedding $\Xi(w)$ is divided into two 3D subspaces as 
$E_6 = E_P \oplus E_D$, where $E_P$ and $E_D$ are the embedding subspaces 
for P- and D-surfaces, respectively. 

It is convenient to define an orthogonal transformation of $E_6$ by the 6D 
orthogonal matrix
\begin{eqnarray}
O(\alpha) := \left(\begin{array}{cr}
{\bf I}\cos\alpha & -{\bf I}\sin\alpha \\
{\bf I}\sin\alpha & {\bf I}\cos\alpha
\end{array}\right)
\end{eqnarray}
with ${\bf I}$ being the 3D identity matrix; the two 3D subspaces, $E_P$ 
and $E_D$, are mixed by such a transformation. Then Eq.(\ref{WEformula}) can 
be written concisely as ${\bf r}(w) = \Pi_P O(\alpha)\Xi(w)$, where 
$\Pi_P$ is the orthogonal projector onto $E_P$ defined by 
$({\bf I}~~{\bf O})$ with ${\bf O}$ being the 3D zero matrix. G-surface is 
obtained by choosing the Bonnet angle as 
$\alpha_G=\arctan{(r/s)}\approx 51.985^{\circ}$. Here the constants $r$ and 
$s$ are defined by $r-is=2 e^{-\pi i/6} K(k^2)$, in which $K(k^2)$ stands 
for the complete elliptic integral of the first kind and $k=e^{-\pi i/3}$. 
Their numerical values are $r\approx 2.156$ and $s\approx 1.686$. 
Hence, G-surface is just the orthogonal projection of $\Xi(w)$ onto a 3D 
subspace $E_G$ which is inclined in the 6D space by the hyper-rotation 
angle $\alpha_G$. The relevant orthogonal projector is given by 
$\Pi_G = \Pi_PO(\alpha_G)$. One may also find that the orthogonal projector 
for D-surface is given by $\Pi_D=\Pi_PO(\alpha_D)=-({\bf O}~~{\bf I})$.

\section{Crystallography of Bonnet transformations}\label{6Ddescription}

With each of the three TPMS's a pair of space groups, $G/H$, is associated; 
$G$ includes the symmetry elements to reverse the sides of the surface, 
while $H$ does not. The group $H$ is a subgroup of $G$. The pairs for P-, 
D-, and G-surfaces are $Im\bar{3}m/Pm\bar{3}m$, $Pn\bar{3}m/Fd\bar{3}m$, 
and $Ia\bar{3}d/I4_132$, respectively. It is important to note that the 
Bravais lattices for P- and D-surfaces depend on whether or not to identify 
the sides of the surfaces. The electronic Hamiltonians for the three TPMS's
to be studied have the symmetry of the group $G$.

\begin{figure}[t]
\begin{center}
\includegraphics[width = 12.5cm]{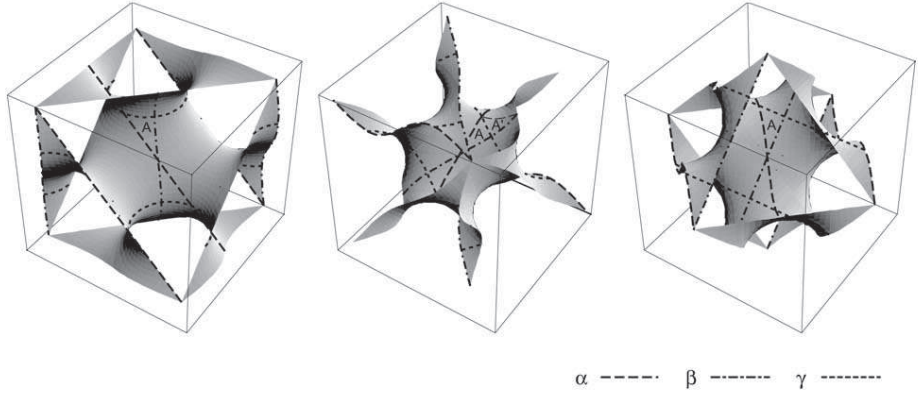}
\caption{The unit cells of the three TPMS's obtained by projecting the 
6D unit cell (corresponding to the dodecagonal region of the 
hyperbolic plane) onto the relevant 3D subspaces $E_P$, $E_D$, and $E_G$. 
Bounding boxes are shown to indicate the relative orientation to the 
cubic unit cell. The fundamental patch consists of either a single 
triangle denoted as {\bf A} (for P- and D-surfaces) or two triangles 
denoted as {\bf A} and {\bf A$^{\prime}$} (for G-surface). 
Representative high-symmetry geodesics of type $\alpha$, $\beta$, and 
$\gamma$ are also indicated on the surfaces. For P-surface these 
correspond to the two-fold axis $U^{\bar{y}z}$ ($\alpha$), the mirror planes 
$\sigma^{xy}$ ($\beta$) and $\sigma^z$ ($\gamma$). For D-surface, 
these are the mirror plane $\sigma^{xy}$ ($\alpha$), the two-fold axes 
$U^{\bar{z}x}$ ($\beta$) and $U^y$ ($\gamma$), respectively. For G-surface, 
no symmetry element corresponds to these geodesics.}\label{UnitPatch}
\end{center}\end{figure}

The WE formulae for the three surfaces reveal that the fundamental patches 
of the surfaces, corresponding to the asymmetric unit of the relevant 
space groups $G$, consist of (2,4,6) triangles.\cite{GCMK00,(pqr)Triangle} 
More precisely, the fundamental patch of P-surface as well as that of 
D-surface coincides with a (2,4,6) triangle. That of G-surface, on the other
hand, consists of two triangles and is twice as large as those of the other 
two (c.f. Fig.\ref{UnitPatch}). The two triangles for the case of G-surface 
are not congruent with each other in the embedding space, so that one of 
them cannot be reproduced via any space group operation from the other. 
However, they are congruent intrinsically, which implies that G-surface 
has an additional ``intrinsic symmetry'' which cannot be reproduced by the 
space group. The surfaces are thus tessellated into (2,4,6) triangles. 

\begin{figure}[htb]
\begin{center}
\includegraphics[width = 8.5cm]{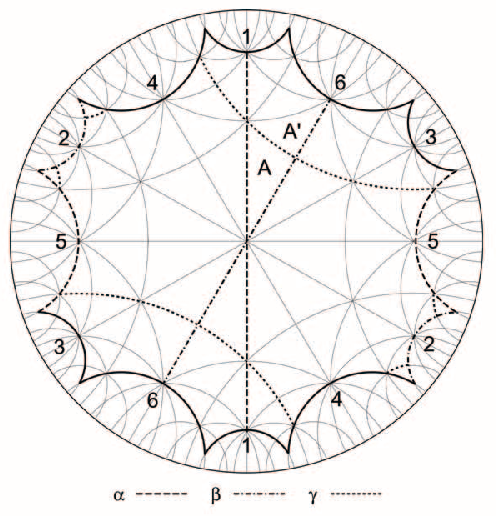}
\caption{The dodecagonal region of the hyperbolic plane, shown within the 
Poincar$\acute{\rm e}$ disc, corresponds to the dodecagonal unit patches 
of P-, D-, and G-surfaces. The identification of the opposite edges yields 
a genus-3 torus corresponding to the unit cell of $\Xi(w)$ with opposite 
sides being identified. The tessellation into (2,4,6)-triangles corresponds 
to the division of the TPMS's into fundamental patches. There are three 
types of geodesics passing through high-symmetry points: $\alpha$, passing 
through six- and four-fold symmetry centers alternately, $\beta$, six- 
and two-fold centers, and $\gamma$ four- and two-fold centers, 
respectively. Only a few representatives of each type is indicated. 
(See also Fig.\ref{UnitPatch}.)}
\label{hyperbolic}
\end{center}\end{figure}

A close relationship between the three TPMS's and the (2,4,6)-tessellation 
of the hyperbolic plane, i.e., the non-Euclid plane with a constant 
Gaussian curvature, was pointed out by Sadoc and Charvolin.\cite{SC89} 
They proved that the translational subgroup of the (2,4,6)-tessellation 
of the hyperbolic plane, where the unit cell is identified as a dodecagonal 
region consisting of 96 triangles as shown in Fig.\ref{hyperbolic}, derives 
the translational subgroups of the three TPMS's. The corresponding Bravais 
lattices are simple cubic (s.c.), face centered cubic (f.c.c.), and body 
centered cubic (b.c.c.) for P-, D-, and G-surfaces, respectively. The unit 
cells of the TPMS's can be taken as dodecagonal patches embedded in three 
dimensions, as shown in Fig.\ref{UnitPatch}, corresponding to the dodecagonal 
region of the hyperbolic plane. Note that these unit cells are only the 
primitive unit cells for the space groups, $H$. If the two sides of a 
surface are indistinguishable, the space group will be $G$ whose primitive 
unit cell is reduced by half for P- and D-surfaces, and the Bravais 
lattices become b.c.c. and s.c., respectively.

\begin{figure}[t]
\begin{center}
\includegraphics[width = 12.5cm]{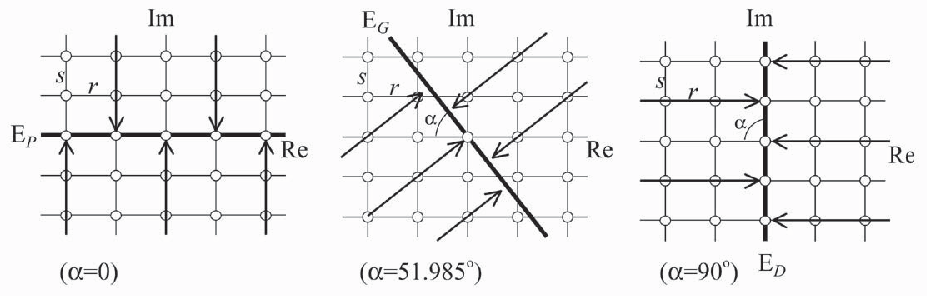}
\caption{The orthogonal projections of the 6D lattice $\Lambda$ for 
P- (left), G- (middle), and D-surfaces (right) are shown schematically 
in the complex plane of each component in ${\bf z}(w)$. The projection 
map of the lattice points of $\Lambda$ are shown by open circles. 
The arrows indicate the directions of the orthogonal projections 
in each plane.}
\label{projection}
\end{center}\end{figure}

The 6D crystallographic description of $\Xi(w)$ was thoroughly investigated 
by Oguey and Sadoc,\cite{OS93} who established that it is a periodic 
structure. The unit cell of the relevant 6D lattice $\Lambda$ is associated 
with the aforementioned dodecagonal patch lifted into the 6D space. The 
primitive lattice vectors of $\Lambda$ are given by the column vectors 
of\cite{OS93}
\begin{eqnarray}
{\bf A}&=&
(\vec{t}_1,\vec{t}_2,\vec{t}_3,\vec{\tau}_1,\vec{\tau}_2,\vec{\tau}_3)
=\ell \left(\begin{array}{cccccc}
r & 0 & 0 & 0 & r & -r \\
0 & r & 0 & -r & 0 & r \\
0 & 0 & r & r & -r & 0 \\
0 & s & -s & 0 & s & s \\
-s & 0 & s & s & 0 & s \\
s & -s & 0 & s & s & 0 \\
\end{array}\right).\label{matrixA}
\end{eqnarray}
The translational symmetry of each of the three surfaces is given as the 3D 
lattice, $\Lambda_X$, with $X=P$, $D$, or $G$. Then it is given as 
$\Lambda_X=\Pi_X\Lambda$. Specifically, $\Lambda_P$ (or $\Lambda_D$) is 
generated by the six column vectors of the upper (or lower) half of 
${\bf A}$, so that it is a s.c. (or f.c.c.) lattice with lattice constant 
$a_P:=\ell r$ (or $a_D:=2 \ell s$). Similarly, $\Lambda_G$ is generated by 
the six column vectors of the matrix
\begin{eqnarray}
\Pi_G {\bf A} &=&
\ell \, b \, \left(\begin{array}{cccccc}
1 & -1 & 1 & 0 & 0 & -2 \\
1 & 1 & -1 & -2 & 0 & 0 \\
-1 & 1 & 1 & 0 & -2 & 0 \\
\end{array}\right)
\end{eqnarray}
with $b=rs/\sqrt{r^2+s^2}$, so that $\Lambda_G$ is a b.c.c. lattice with 
lattice constant $a_G:=2 \ell b$. In each case, the unit cell is given by 
the projection of the dodecagonal patch onto the relevant subspace 
(see Fig.\ref{UnitPatch}). The relationship between $\Lambda$ and $\Lambda_X$
($X=P$, $D$, and $G$) is schematically represented in Fig.\ref{projection}.

Let us consider the reciprocal lattice $\Lambda^{\ast}$ of $\Lambda$. 
To do so, we shall take the transposed inverse of {\bf A}, 
${\bf B} = \left({\bf A}^{-1}\right)^{T}$, which is given as
\begin{eqnarray}
{\bf B} = \frac{1}{2\ell}
\left(\begin{array}{cccccc}
0 & \bar{r} & \bar{r} & 0 & \bar{r} & -\bar{r} \\
\bar{r} & 0 & \bar{r} & -\bar{r} & 0 & \bar{r} \\
\bar{r} & \bar{r} & 0 & \bar{r} & -\bar{r} & 0 \\
0 & \bar{s} & -\bar{s} & \bar{s} & 0 & 0 \\
-\bar{s} & 0 & \bar{s} & 0 & \bar{s} & 0 \\
\bar{s} & -\bar{s} & 0 & 0 & 0 & \bar{s} \\
\end{array}\right),
\end{eqnarray}
with $\bar{r}=1/r$ and $\bar{s}=1/s$. The generators of $\Lambda^{\ast}$ 
are given as the column vectors of ${\bf B}$.
We shall denote these generators as $\vec{Q}_i$ ($i=1,...,6$), so that 
${\bf B}=(\vec{Q}_1,\vec{Q}_2,\vec{Q}_3,\vec{Q}_4,\vec{Q}_5,\vec{Q}_6)$.

Since the TPMS's are projections of $\Xi(w)$ onto the 3D subspaces, 
$E_P$, $E_D$, and $E_G$, their reciprocal lattices, $\Lambda_P^{\ast}$, 
$\Lambda_D^{\ast}$, and $\Lambda_G^{\ast}$, are the intersections of 
$\Lambda^{\ast}$ along the relevant 3D subspaces. It can be shown that the 
sets of the three generators of these reciprocal lattices are given by
\begin{eqnarray}
\begin{array}{ll}
\left\{~\vec{Q_1}+\vec{Q_5}-\vec{Q_6},~
       \vec{Q_2}-\vec{Q_4}+\vec{Q_6},~
       \vec{Q_3}+\vec{Q_4}-\vec{Q_5}~\right\} \quad
     &  {\rm for} ~ \Lambda_P^{\ast},\\
\left\{~-\vec{Q_2}+\vec{Q_3}+\vec{Q_4},~
       \vec{Q_1}-\vec{Q_3}+\vec{Q_5},~
       -\vec{Q_1}+\vec{Q_2}+\vec{Q_6}~\right\} \quad
     &  {\rm for} ~ \Lambda_D^{\ast},\\
\left\{~\vec{Q_1}-\vec{Q_4}-\vec{Q_6},~
       \vec{Q_2}-\vec{Q_4}-\vec{Q_5},~
       \vec{Q_3}-\vec{Q_5}-\vec{Q_6}~\right\} \quad
     &  {\rm for} ~ \Lambda_G^{\ast}.
\end{array} \label{generatorLast}
\end{eqnarray}

\section{Electronic band structures of the TPMS's}\label{bandstruc}

For clarity, we shall focus on a specific situation in which the electron 
is constrained to move along the surface, which means that the electron 
cannot propagate through the region outside the surface. In order to 
realize the situation, an infinite potential well is introduced to confine 
the electron within a thin layer of constant thickness $d$ built over the 
surface. Then by taking the limit of $d \to 0^{+}$, one obtains the 
Schr$\ddot{\rm o}$dinger equation for the particle constrained to move 
along the surface.\cite{JK71,C81,C82,IN91} Let us assume a curvilinear 
coordinates $(q^1,q^2)$ on the surface, for which the metric tensor 
$g_{ij}$ ($i,j=1,2$) is defined by 
${\rm d} s^2 = \sum_{ij} g_{ij}{\rm d} q^i {\rm d} q^j$ where ${\rm d} s$ 
is an infinitesimal distance. The principal curvatures are denoted as 
$\kappa_1=1/R_1$ and $\kappa_2=1/R_2$, with $R_1$ and $R_2$ being the 
radii of curvature. The Schr$\ddot{\rm o}$dinger equation is written as,
\begin{eqnarray}\label{SchEq}
-\frac{\hbar^2}{2m}\frac{1}{\sqrt{g}}\sum_{i,j=1}^2
\frac{\partial}{\partial q^i}
\sqrt{g}g^{ij}\frac{\partial}{\partial q^j}\psi(q^1,q^2)
-\frac{\hbar^2}{8m}(\kappa_1-\kappa_2)^2\psi(q^1,q^2)=E\psi(q^1,q^2),
\end{eqnarray}
where $g={\det{(g_{ij})}}$ and $(g^{ij})=(g_{ij})^{-1}$. The first term 
describes the propagation of the wave function along the curved surface. 
The second term is the attractive potential energy associated with the 
local curvature,\cite{vsclassical} resulting from the quantization 
procedure of the motion perpendicular to the surface. The normalization 
integral for $\psi$ is given by
\begin{eqnarray}
\int\int |\psi(q^1,q^2)|^2 \sqrt{g} {\rm d} q^1 {\rm d} q^2,
\end{eqnarray}
which is set to unity for square integrable wave functions.

The above model, though seemingly too idealistic, is useful in answering 
fundamental questions about the quantum effects associated with geometrical 
surfaces. It may actually suit nano-structured materials composed of 
conductive thin curved films possibly to be realized in a near future. 
Since the model ignores the individual atomic potential energies, suitable 
systems should consist of substances with large Fermi wave length compared 
to the surface thickness or atomic distances, which may be achieved by 
semi-conducting substances (including graphene sheets).

In order to study the electronic structures of the TPMS's, the WE 
representation must be incorporated to the above equation. Then one can 
study the relationship between the geometry, that is, the symmetry and 
topology, of the TPMS's and the electronic properties. This may be a useful 
approach to understand fundamental aspects of the problem for more general 
cases than minimal surfaces. Aoki et al.\cite{AKTMK01} formulated the 
Schr$\ddot{\rm o}$dinger equation with the stereographic projection, 
$w=\cot{(\theta/2)}\exp{(i\phi)}$, into the following form:
\begin{eqnarray}
-\frac{(1-\cos{\theta})^4}{\left|F(w)\right|^2}\left(
\frac{\partial^2}{\partial \theta^2}+
\cot{\theta}\frac{\partial}{\partial \theta}+
\frac{1}{\sin^2{\theta}}\frac{\partial^2}{\partial\phi^2}+1
\right)\psi(\theta,\phi)= \epsilon \psi(\theta,\phi).
\end{eqnarray}
Here the energy eigenvalue is represented by the dimensionless variable 
$\epsilon=8m\ell^2E/\hbar^2$. $\epsilon$ is a universal quantity among 
surfaces which just differ by scale, whereas the unit of energy, 
$\hbar^2/(8m\ell^2)$, depends on the scale (it is 9.525 meV for 
$\ell=1.0$ nm, for instance). Remarkably, the above equation is free from 
the phase factor $e^{i\alpha}$ ($\alpha$ is the Bonnet angle), so that 
it is identical for P-, G-, and D-surfaces. This is a result of the 
isometry of these surfaces as well as that the curvature potential depends 
only on the intrinsic (i.e., Gaussian) curvature for minimal surfaces, since
$-\frac{\hbar^2}{8m}(\kappa_1-\kappa_2)^2=\frac{\hbar^2}{2m}\kappa_1\kappa_2$.
Here we have used the vanishing of mean curvature, $\kappa_1+\kappa_2=0$.

The topological differences among the surfaces should be taken into account 
through suitable choices of boundary conditions on the unit cell. Since the 
topology is crucial in restricting the propagation and interference of 
electronic wave functions, the electronic structures should largely depend 
on it. It is an interesting question to elucidate how the band structures 
are affected by the topology.

\begin{figure}[htb]
\begin{center}
\includegraphics[width = 11cm]{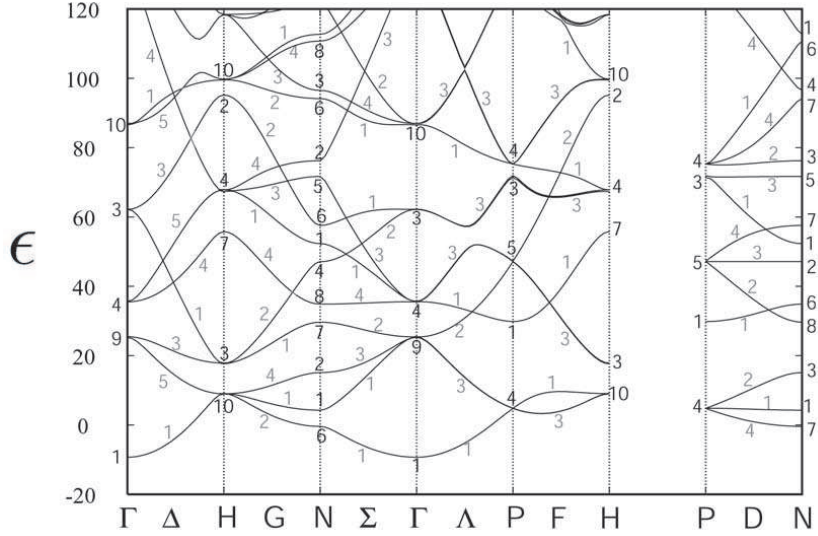}
\caption{The band structures of P-surface. The numbers shown on each of 
the energy levels indicate the corresponding IR's following the numbering 
scheme of Ref.\onlinecite{ZakTables}. The numbers are shown in black for 
special points and in gray for special lines.}\label{Pband_new}
\end{center}\end{figure}

\begin{figure}[htb]
\begin{center}
\includegraphics[width = 11cm]{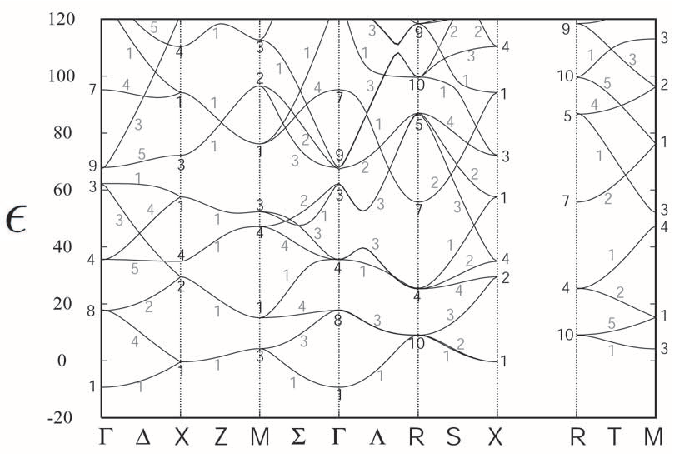}
\caption{The band structures of D-surface. See the figure caption of 
Fig.\ref{Pband_new}.}\label{Dband_new}
\end{center}\end{figure}

\begin{figure}[htb]
\begin{center}
\includegraphics[width = 11cm]{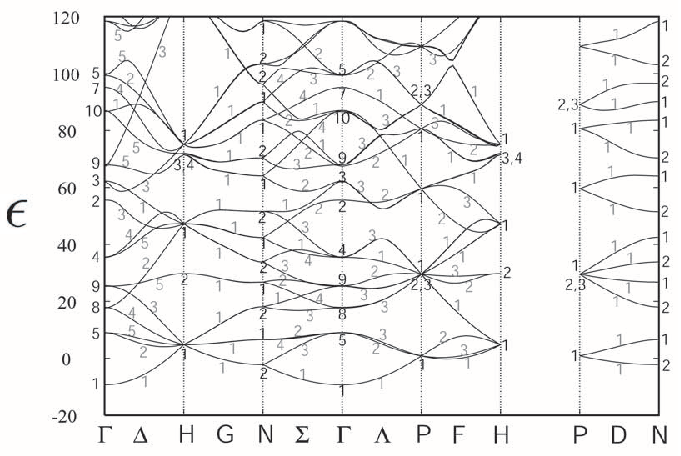}
\caption{The band structures of G-surface. See the figure caption of 
Fig.\ref{Pband_new}.}\label{Gband_new}
\end{center}\end{figure}

The Schr$\ddot{\rm o}$dinger equation can be solved numerically, by 
discretizing the coordinates $\theta$ and $\phi$ and applying the finite 
difference method. In choosing the mesh values, special care is needed in 
the vicinity of the singularity at the branch points of the WE integrand 
to avoid the degradation of the resulting precision.\cite{AKTMK01} We 
employ a non-uniform mesh in which the maximal interval in the 3D embedding 
space would be minimized for a fixed number of mesh points. We calculate 
the band structures for P-, G-, and D-surfaces by using 4096 mesh points 
per dodecagonal unit of Fig.\ref{UnitPatch}. The results are shown in 
Figs.\ref{Pband_new}, \ref{Dband_new}, and \ref{Gband_new}.

\section{Interrelation between the band structures}\label{6Dband}

The crystallographic information given in Sec.\ref{6Ddescription} provides 
a basis for understanding the interrelations between the electronic 
structures of these TPMS's. The boundary conditions for a dodecagonal 
patch of each surface is given as the phase factor multiplied upon 
translating it to any of its neighbors. The phase factor is given by 
$e^{2\pi i {\bf k}\cdot{\bf t}}$ for the translational vector ${\bf t}$, 
where ${\bf k}$ is the 3D wave vector. Instead of the 3D vectors ${\bf k}$ 
and ${\bf t}$, one can use the corresponding 6D vectors $\vec{k}$ and 
$\vec{t}$ which are restricted within the relevant 3D subspace. The 
phase factor is then rewritten as $e^{2\pi i \vec{k}\cdot\vec{t}}$, in 
which the vector $\vec{t}$ can safely be lifted to a 6D translational 
vector of $\Xi(w)$ prior to the orthogonal projection. The boundary 
conditions can be generalized further by allowing the vector $\vec{k}$, 
which should otherwise be restricted within the relevant 3D subspace for 
each surface, to take arbitrary values in the hypothetical 6D 
``reciprocal space''. Thus we are led to define the 6D band structures, 
$\hat{\epsilon}(\vec{k})$, which are periodic functions of the 
quasi-momentum, $\vec{k}=(k_1,k_2,k_3,k_4,k_5,k_6)$. 
Since $\hat{\epsilon}(\vec{k})$ are derived from the periodicity of 
$\Xi(w)$, they are periodic functions of $\vec{k}$ with the periodicity 
of $\Lambda^{\ast}$. It is understood that the band structures for each 
of the three TPMS's are the section of $\hat{\epsilon}(\vec{k})$ along the 
relevant 3D subspace. Hence, if $\vec{k}_X$ is an arbitrary 6D vector 
along $E_X$ ($X=P,D,$ or $G$) and ${\bf k}_X$ is the 3D counterpart, 
then the band structures for X-surface is given as 
$\epsilon_X({\bf k}_X)=\hat{\epsilon}(\vec{k}_X)$.

It is readily shown that the intersection of a pair of subspaces, $E_X$ 
and $E_Y$ ($X, Y = P, D,$ or $G$), consists of just one point, which is 
at the origin. As a consequence, the energy eigenvalues of X-surface at 
$\Gamma$-point always coincide with those of Y-surface at $\Gamma$-point, 
because $\epsilon_X({\bf 0})=\epsilon_Y({\bf 0})=\hat{\epsilon}(\vec{0})$. 
In this case, we can say that the band structures of X- and Y-surfaces 
``intersect'' at $\Gamma$-point.

A more general condition for the band intersection is given below. Recall 
$\hat{\epsilon}(\vec{k})$ has the periodicity of $\Lambda^{\ast}$, so that
\begin{eqnarray}
\hat{\epsilon}(\vec{k}+\sum_i n_i \vec{Q}_i)=\hat{\epsilon}(\vec{k}),
\end{eqnarray}
in which $n_i$ ($i=1,...,6$) are arbitrary integer coefficients. 
Therefore, if the two vectors $\vec{k}_X$ in $E_X$ and $\vec{k}_Y$ in 
$E_Y$ are connected through a scattering vector $\sum_i n_i \vec{Q}_i$ 
in $\Lambda^{\ast}$,
\begin{eqnarray}\label{bandintersection}
\vec{k}_Y=\vec{k}_X+\sum_i n_i \vec{Q}_i,
\end{eqnarray}
then the relevant eigenvalues coincide, i.e., 
$\epsilon_X({\bf k}_X)=\epsilon_Y({\bf k}_Y)$. In particular, if 
$\sum_i n_i\vec{Q}_i \in \Lambda^{\ast}_X\oplus \Lambda^{\ast}_Y$, then 
one immediately finds that $\vec{k}_X \in \Lambda^{\ast}_X$ and 
$\vec{k}_Y \in \Lambda^{\ast}_Y$. Since these wave vectors are at the 
center of the Brillouin zones (BZ's), the band intersection corresponds 
to $\Gamma$-point as discussed earlier. On the other hand, if 
$\sum_i n_i\vec{Q}_i$ does not belong to 
$\Lambda^{\ast}_X\oplus \Lambda^{\ast}_Y$, it is divided into two 
components in $E_X$ and $E_Y$ which are not at the center of the BZ's. 
The latter case leads to band intersections at certain special points in 
the BZ's. The total number of band intersections within the 6D BZ is given 
as the index $\mu$ of the sublattice $\Lambda^{\ast}_X\oplus\Lambda^{\ast}_Y$
in $\Lambda^{\ast}$.\cite{SublatticeIndex} It can be shown that $\mu=4$ 
for the pair of P- and D-surfaces, 2 for that of P- and G-surfaces, 
and 1 for that of D- and G-surfaces. The corresponding band intersections 
occur at a set of special points in the BZ's of these TPMS's, as listed 
in TABLE \ref{T1}. (For the derivation, see Appendix \ref{Appendix:Indices}.) 
\begin{table}[t]
\begin{center}\begin{tabular}{ccc}
\hline\hline
\makebox[3cm]{P-surface}&\makebox[3cm]{D-surface}&\makebox[3cm]{G-surface} \\
\hline
$\Gamma$ ($\Gamma$,$H$) & $\Gamma$ ($\Gamma$,$R$) & $\Gamma$ \\
$M$ $(N^2)$             & $X$ $(M,X)$             &          \\
$R$ $(P^2)$             &                         & $H$      \\
\hline\hline
\end{tabular}\end{center}
\caption{The list of the special points for the band intersections, 
as derived in Appendix \ref{Appendix:Indices}. The special points for 
the Bravais lattices of the dodecagonal unit patches are shown without 
parenthesis. The corresponding special points in the unfolded BZ's 
for P- and D-surfaces are shown by parenthesis. A superscript is added 
if two special points in the same star correspond to a single band 
intersection.}\label{T1}
\end{table}

\begin{figure}[htb]
\begin{center}
\includegraphics[width = 9.5cm]{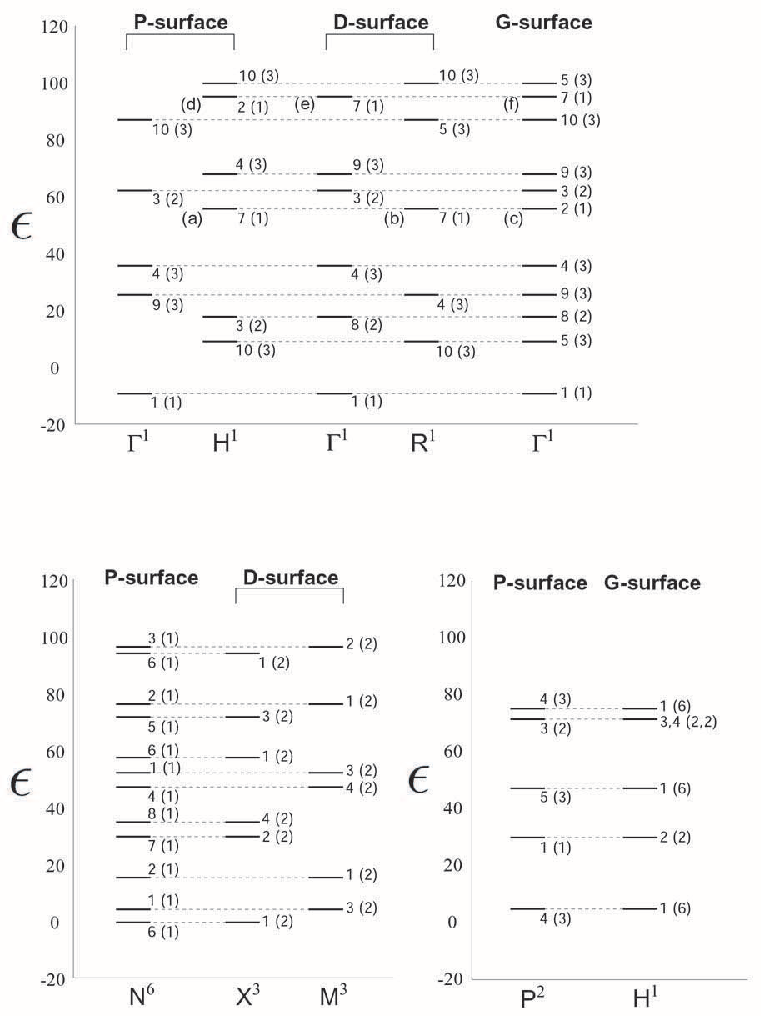}
\caption{Energy diagrams showing the coincidence of energy levels at the 
special points listed in TABLE \ref{T1}. The superscript for each special 
point symbol (e.g., N$^6$) indicates the number of the equivalent points 
in the BZ, constituting the star. Each energy level is assigned the 
relevant IR, whose dimensionality is shown by parenthesis.}
\label{BandIntersections2}
\end{center}\end{figure}

The coincidence of energy eigenvalues as the result of band intersections 
are observed in our numerical results as demonstrated in 
Fig.\ref{BandIntersections2}. This phenomenon has been found previously
by Aoki et al.\cite{AKTMK01,KA05} in an empirical way. However, our theory 
provides a more clear account for this phenomenon in terms of the 6D 
algebra of Bonnet transformations. It also gives the exhaustive list 
of possible band intersections in TABLE \ref{T1}. Moreover, the idea of 
band intersections can be applied generally to the band structures of 
TPMS's connected via Bonnet transformations.

\section{Symmetry analysis of the band structures}\label{symmetry}

Symmetry has various implications in understanding the properties of 
materials involving the quantum mechanics of electrons. Dynamically, it 
implies the selection rules for quantum transitions or various forms of 
conservation laws. While statically, symmetry serves as the classification 
principle of structures and also provides basic clues for understanding 
the formation (or stability) of a particular material. It should also have
significant implications in understanding the detailed characteristics of 
the present systems. Therefore, in the following, we shall carry out a 
detailed analysis on the symmetry aspects of the electronic structures of
the TPMS's.

A subspace of Bloch states $\phi_{\bf k}$ at a wave vector, ${\bf k}$, 
with an identical energy eigenvalue can be assigned an irreducible 
representation (IR) of $G_{\bf k}$, i.e., the group of ${\bf k}$. 
In practice, the representation matrices are obtained numerically, and 
are then analyzed by the use of the orthogonal relationships of 
IR's.\cite{Lax,TSPACE} In fact, the discretization of the surfaces used 
for the numerical analysis\cite{AKTMK01} diminishes the space group, $G$, 
to $P4/mmm$, $P4_2/nnm$, and $I4_1/acd$ for P-, D-, and G-surfaces, 
respectively. Hence, for each surface, we can numerically analyze the IR's 
only at ${\bf k}$-points whose group $G_{\bf k}$ are subgroups of the 
diminished space group. In addition, we may use the band branching rules 
(for instance, $\Gamma_9$ always splits into $\Delta_3$ and $\Delta_5$ for 
the space group $Im\bar{3}m$) to determine the IR's for the rest of the 
${\bf k}$ space. The numbers indicated on the band structure functions 
in Figs.\ref{Pband_new}, \ref{Dband_new}, and \ref{Gband_new} represent 
the relevant IR's of $G_{\bf k}$; we comply with the numbering scheme 
given in Ref.\onlinecite{ZakTables}. These IR's describe the symmetry 
features of the relevant eigenstates. They also serve as criteria for the 
degeneracies in the band structures. This is useful since otherwise 
degenerate bands can slightly fall apart as artifacts of the diminishing 
of symmetry involved by the discretization procedure.

\subsection{Band sticking}

Degeneracies occur due to several symmetry reasons. At high-symmetry 
special points in the BZ, we often observe several band energy curves 
sticking together into one energy level. The degeneracy of such an energy 
level is given by the dimensionality of the relevant IR. For the case of 
non-symmorphic space groups, the degeneracies at a special point on the 
boundary of the BZ can be larger than symmorphic counterparts when the 
transformation of eigenstates by $G_{\bf k}$ is represented as a multiplier 
group.\cite{Lax} In the band structures of G-surface, for instance, we may 
observe relatively large degeneracies at H- and P-points; they are 6-fold 
(H$_1$), 4-fold (P$_1$), and 2-fold (H$_2$, H$_3$, H$_4$, P$_2$, P$_3$). 
Furthermore, we may find additional degeneracies due to the time-reversal 
symmetry on these special points.\cite{Herring,Lax} Namely, an energy 
level of the IR H$_3$ always sticks with one of the IR H$_4$, whereas the 
same holds for the pair of the IR's, P$_2$ and P$_3$, both resulting in 
4-fold degeneracies. Note the space group for G-surface, $Ia\bar{3}d$, is 
highly non-symmorphic including four-fold screw axes and two types of glide 
planes (one type is d-glide).

\subsection{Nodal lines}

An interesting feature of the present systems is that, due to the 2D 
character of the surfaces, some of the eigenstates have nodal lines, i.e., 
lines along which an eigenstate has zero amplitude. Here, we shall study 
the occurrence of nodal lines in somewhat detail from the symmetry 
standpoint, because nodal lines have the following significant implications 
for the electronic properties. Firstly, the density of nodal lines is 
correlated with the magnitude of energy eigenvalue. Secondly, nodal lines 
occur as a consequence of the quantum interference of an electronic wave 
and are closely related to the symmetry and topology of the surfaces. 

There is an important class of nodal lines which can be understood in terms 
of the symmetry properties of the eigenstates; a nodal line should appear 
when the relevant IR has odd parity with respect to a symmetry element 
which leaves the positions of the nodal line invariant. There are two types 
of such symmetry elements: mirrors (a nodal line will appear at the 
intersection between the surface and a mirror plane) and two-fold axes 
lying in the surface (a nodal line will appear along such an axis).

In the space group $Im\bar{3}m$ of P-surface, the symmetry elements 
related to the generation of nodal lines are
\begin{itemize}
\item (two-fold axes $U_d$) $U^{xy}$, $U^{\bar{x}y}$, $U^{yz}$, 
$U^{\bar{y}z}$, $U^{zx}$, $U^{\bar{z}x}$,
\item (mirrors $\sigma_d$) $\sigma^{xy}$, $\sigma^{\bar{x}y}$, 
$\sigma^{yz}$, $\sigma^{\bar{y}z}$, $\sigma^{zx}$, $\sigma^{\bar{z}x}$,
\item (mirrors $\sigma$) $\sigma^x$, $\sigma^y$, $\sigma^z$.
\end{itemize}
Note that the elements of the factor group are implied by the symbols of
point group elements. Throughout this paper, the symbols of point 
group elements comply with those used in Ref.\onlinecite{ZakTables}. 
One can find that the above three types of symmetry operations leave the 
high-symmetry geodesics on P-surface, i.e., $\alpha$-, $\beta$-, and 
$\gamma$-type geodesics as defined in Fig.\ref{hyperbolic}, invariant. 
The correspondence between the types of symmetry operations and those of 
geodesics are given by $\alpha\leftrightarrow U_d$, 
$\beta\leftrightarrow\sigma_d$, and 
$\gamma\leftrightarrow\sigma$ (see Fig.\ref{UnitPatch}). 

A similar list for D-surface (space group $Pn\bar{3}m$) is given as
\begin{itemize}
\item (mirrors $\sigma_d$) $\sigma^{xy}$, $\sigma^{\bar{x}y}$, $\sigma^{yz}$, $\sigma^{\bar{y}z}$, $\sigma^{zx}$, $\sigma^{\bar{z}x}$,
\item (two-fold axes $U_d$) $U^{xy}$, $U^{\bar{x}y}$, $U^{yz}$, $U^{\bar{y}z}$, $U^{zx}$, $U^{\bar{z}x}$,
\item (two-fold axes $U$) $U^{x}$, $U^{y}$, $U^{z}$.
\end{itemize}
In this case the correspondence is established as, 
$\alpha\leftrightarrow\sigma_d$, $\beta\leftrightarrow U_d$, and 
$\gamma\leftrightarrow U$ (see Fig.\ref{UnitPatch}). 

On the other hand, no such symmetry elements exist for G-surface, that is, 
there are no mirror planes (though there are several glide planes) nor 
two-fold axes lying in the surface. Hence no nodal lines are predicted 
by symmetry for G-surface.

It is deducible from the character table of a given ${\bf k}$-point in the 
BZ whether a given IR entails any nodal lines by symmetry. One can consult 
the character tables for the space groups in Ref.\onlinecite{ZakTables}. 
It is necessary from the previous paragraph that the group of ${\bf k}$, 
$G_{\bf k}$, contains at least one symmetry element(s) responsible for the 
generation of nodal lines. Then, if an IR of $G_{\bf k}$ has odd parity 
with respect to those symmetry elements, a nodal line will appear 
accordingly. One should bear in mind that the representative of the relevant 
factor group $(\gamma |{\bf c})$ ($\gamma$ is the point group operation and 
${\bf c}$ the additional translation, which is usually taken to be zero 
when the space group is symmorphic) does not necessarily leave the 
position of the nodal line invariant. Instead, one should consider 
$(\gamma |{\bf r})$ with
\begin{eqnarray}
{\bf r}={\bf q}-\gamma{\bf q}={\bf c}-{\bf R}_{\bf q}^{(\gamma |{\bf c})},
\end{eqnarray}
where ${\bf q}$ is any position in the relevant mirror or axis. 
Here, ${\bf R}_{\bf q}^{(\gamma |{\bf c})}=(\gamma |{\bf c}){\bf q}-{\bf q}$
is always a Bravais vector. The relevant matrix for 
$(\gamma |{\bf r})$ is given by
\begin{eqnarray}
D[(\gamma |{\bf r})]=\exp{ [-i{\bf k R}_{\bf q}^{(\gamma |{\bf c})}] }
D[(\gamma |{\bf c})].
\end{eqnarray}
Therefore, in order to discern the parity of an IR at the position of a 
nodal line, the character for the representative element of the relevant
factor group (the value given in the character table) should be multiplied 
by $\exp{ [-i{\bf k R}_{\bf q}^{(\gamma |{\bf c})}] }$. 
The exhaustive lists of IR's which entail nodal lines by the above argument
are given in TABLES \ref{T2} and \ref{T3} for P- and D-surfaces, respectively.
For each of the IR's listed, the corresponding nodal lines are given by
parenthesis.

\begin{table}[t]
\begin{center}\begin{tabular}{ccc}
\hline\hline
${\bf k}$-point & $G_k/T$ & IR (nodal lines)  \\
\hline
$\Gamma~(0,0,0)$ & $O_h$ & 
$\Gamma_2(6U_d,6\sigma_d),\Gamma_6(3\sigma,6\sigma_d),
\Gamma_7(6U_d,3\sigma),\Gamma_8(3\sigma)$ \\
$\Delta~(k_x,0,0)$ & $C_{4v}$ & 
$\Delta_2(2\sigma,2\sigma_d),\Delta_3(2\sigma_d),\Delta_4(2\sigma)$ \\
$H~(1,0,0)$ & $O_h$ & 
$H_1(6U_d),H_2(6\sigma_d),H_6(6U_d,3\sigma,6\sigma_d),
H_7(3\sigma),H_8(3\sigma)$ \\
$\Sigma~(k_x,k_x,0)$ & $C_{2v}$ & 
$\Sigma_2(U_d,\sigma_d),\Sigma_3(\sigma,\sigma_d),\Sigma_4(U_d,\sigma)$ \\
$G~(k_x,1-k_x,0)$ & $C_{2v}$ & 
$G_1(U_d),G_2(\sigma_d),G_3(U_d,\sigma,\sigma_d),G_4(\sigma)$ \\
$N~(\frac{1}{2},\frac{1}{2},0)$ & $D_{2h}$ & 
$N_1(U^{\bar{x}y}),N_2(\sigma^{\bar{x}y},\sigma^z),
N_3(U^{xy},U^{\bar{x}y},\sigma^z,\sigma^{xy}),
N_4(U^{xy},\sigma^{xy},\sigma^{\bar{x}y}),$\\
{} & {} & $N_5(U^{\bar{x}y},\sigma^{z},\sigma^{xy},\sigma^{\bar{x}y}),
N_6(\sigma^{xy}),N_7(U^{xy},U^{\bar{x}y},\sigma^{\bar{x}y}),
N_8(U^{xy},\sigma^z)$ \\
$\Lambda~(k_x,k_x,k_x)$ & $C_{3v}$ & $\Lambda_2(3\sigma_d)$ \\
$F~(k_x,1-k_x,1-k_x)$ & $C_{3v}$ & $F_2(3\sigma_d)$ \\
$D~(\frac{1}{2},\frac{1}{2},k_z)$ & $C_{2v}$ & 
$D_2(\sigma^{xy}),D_3(\sigma^{xy},\sigma^{\bar{x}y}),D_4(\sigma^{\bar{x}y})$ \\
$P~(\frac{1}{2},\frac{1}{2},\frac{1}{2})$ & $T_d$ & $P_2(6\sigma_d)$ \\
$\Xi~(k_x,k_y,0)$ & $C_s$ & $\Xi_2(\sigma)$ \\
$\Theta~(k_x,k_x,k_z)$ & $C_s$ & $\Theta_2(\sigma_d)$ \\
$B~(k_x,1-k_x,k_z)$ & $C_s$ & $B_2(\sigma_d)$ \\
\hline\hline
\end{tabular}\end{center}
\caption{The list of the IR's for P-surface ($Im\bar{3}m$) entailing 
nodal lines by symmetry. The IR's shown have odd parity with respect to 
the symmetry elements given by parenthesis; these elements are of the 
types, $U_d$, $\sigma_d$, and $\sigma$, which are responsible for the 
generation of nodal lines along $\alpha$-, $\beta$-, and 
$\gamma$-geodesics, respectively. A set of symmetrically equivalent 
elements are shown by the numbered type symbols; e.g., 
$3\sigma=\{\sigma_x,\sigma_y,\sigma_y \}$.}\label{T2}
\end{table}

\begin{table}[t]
\begin{center}\begin{tabular}{ccc}
\hline\hline
${\bf k}$-point & $G_k/T$ & IR (nodal lines)  \\
\hline
$\Gamma~(0,0,0)$ & $O_h$ & 
$\Gamma_2(6U_d,6\sigma_d),\Gamma_6(6\sigma_d),\Gamma_7(6U_d)$ \\
$\Delta~(k_x,0,0)$ & $C_{4v}$ & 
$\Delta_2(2\sigma_d),\Delta_3(2\sigma_d),\Delta_5(U)$ \\
$X~(\frac{1}{2},0,0)$ & $D_{4h}$ & 
$X_2(2\sigma_d),X_3(U^x,U^{yz},U^{\bar{y}z}),X_4(U^x)$ \\
$\Sigma~(k_x,k_x,0)$ & $C_{2v}$ & 
$\Sigma_2(\sigma_d,U_d),\Sigma_3(\sigma_d),\Sigma_4(U_d)$ \\
$Z~(k_x,\frac{1}{2},0)$ & $C_{2v}$ & {---} \\
$M~(\frac{1}{2},\frac{1}{2},0)$ & $D_{4h}$ & 
$M_1(U^z),M_2(U^z,2\sigma_d),M_4(U^{xy},U^{\bar{x}y})$ \\
$\Lambda~(k_x,k_x,k_x)$ & $C_{3v}$ & $\Lambda_2(3\sigma_d)$ \\
$S~(\frac{1}{2},k_y,k_y)$ & $C_{2v}$ & 
$S_3(\sigma^{\bar{y}z}),S_4(\sigma^{\bar{y}z})$ \\
$T~(\frac{1}{2},\frac{1}{2},k_z)$ & $C_{4v}$ & 
$T_2(U^z),T_3(U^z),T_4(U^z),T_5(U^z)$ \\
$R~(\frac{1}{2},\frac{1}{2},\frac{1}{2})$ & $O_h$ & 
$R_1(3U,6U_d),R_2(3U,6\sigma_d),R_3(3U),R_6(3U,6U_d,6\sigma_d),
R_7(3U),R_8(3U)$ \\
$\Xi~(k_x,k_y,0)$ & $C_s$ & {---} \\
$\Theta~(k_x,k_x,k_z)$ & $C_s$ & $\Theta_2(\sigma_d)$ \\
$A~(\frac{1}{2},k_y,k_z)$ & $C_s$ & {---} \\
\hline\hline
\end{tabular}\end{center}
\caption{The list of the IR's for D-surface ($Pn\bar{3}m$) entailing 
nodal lines by symmetry. The IR's shown have odd parity with respect to 
the symmetry elements of the types, $\sigma_d$, $U_d$, and $U$, which are 
responsible for the generation of nodal lines along 
$\alpha$-, $\beta$-, and $\gamma$-geodesics, respectively.}\label{T3}
\end{table}

\begin{figure}[htb]
\begin{center}
\includegraphics[width = 9.5cm]{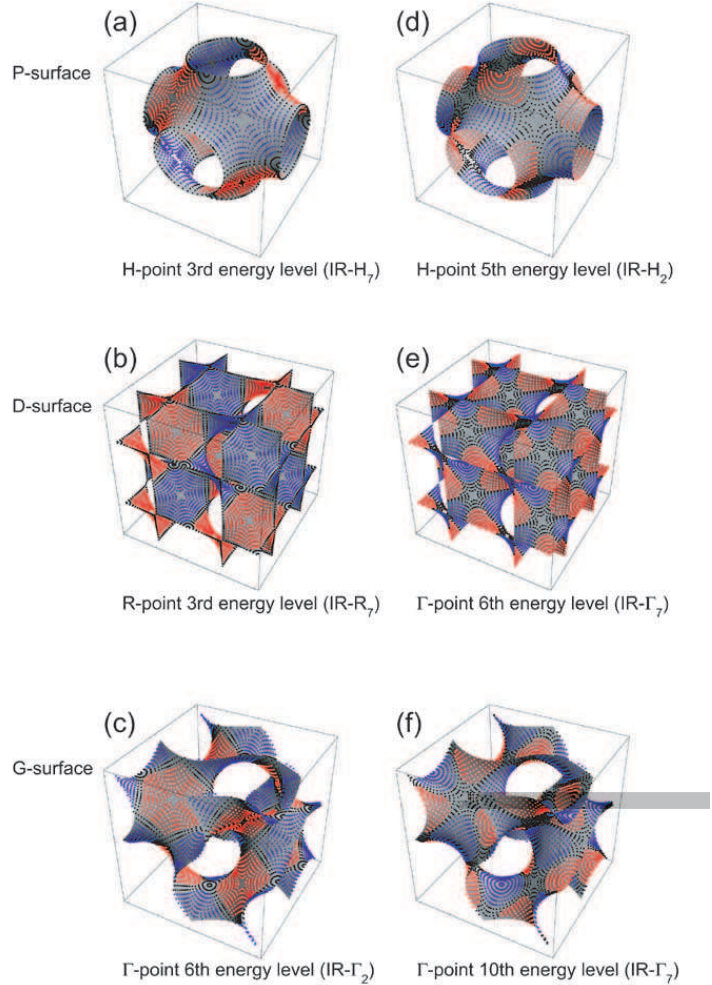}
\caption{(Color) In each column eigenstates with identical energy 
eigenvalue corresponding to a band intersection are shown. The dots are 
the actual mesh points for the numerical calculation. Opposite signs of the 
wave functions are indicated by the blue and red colors, while the black 
dots indicate the amplitude whose absolute value is close to zero, 
corresponding to the nodal lines. The states in the set \{(a), (b), (c)\} 
have nodal lines along all the $\gamma$ type geodesics, while ones in 
the set \{(d), (e), (f)\} have nodal lines along all the $\beta$ type 
geodesics. The $\gamma$ type geodesics for G-surface in (c) constitute 
the helical lines of both chirality along $x$, $y$, and $z$ directions .}
\label{wavefunc}
\end{center}\end{figure}

For G-surface, we may still find related nodal lines for eigenstates 
on the special points $\Gamma$ and $H$, corresponding to band intersections. 
These eigenstates are in one-to-one correspondence with eigenstates of P- 
and D-surfaces on the relevant special points (see also a discussion in
Sec.\ref{discussions}),
and they are identical intrinsically within the dodecagonal unit patch. 
Therefore if some nodal lines are present for symmetry reasons for P- and 
D-surfaces, the corresponding states of G-surface should also have nodal 
lines. Illustrative examples are given in Fig.\ref{wavefunc}, which shows 
two sets of eigenstates with IR's \{$H_7$, $R_7$, $\Gamma_2$\} and \{$H_2$, 
$\Gamma_7$, $\Gamma_7$\} for \{P, D, G\}-surfaces, respectively. One can see
that the sets of nodal lines are intrinsically the same within each set.

We find that some eigenstates have nodal lines which do not correspond to 
any of the three types of high symmetry geodesics. This kind of nodal 
lines are by no means easy to predict because they are {\em not} directly 
related to mirrors nor two-fold axes in the surface. However, such nodal 
lines would appear only at sufficiently high energies.

\section{Discussions}\label{discussions}

The idea of band intersections discussed in Sec.\ref{6Dband} can be 
paraphrased in the following way. The Bloch condition for the eigenstates 
of a crystal is given as the proper phase relationships of the reference 
unit cell to its neighbors. For the present case, we have shown that the 
phase relationships of a dodecagonal unit to its twelve neighbors become 
identical between two different surfaces at the {\bf k} points of a band 
intersection. Therefore, the partial Hamiltonians $h_{\bf k}$ acting on 
the subspaces of states at these {\bf k} points will become equivalent. 
The coincidence of energy eigenvalues is just a result of this fact. 
If there are several members in the star of the special points, all these 
members give a set of symmetrically equivalent phase relationships of
the dodecagonal unit patch; for instance, the stars of M-point (s.c.) of 
P-surface and X-point (f.c.c.) of D-surface forming band intersections are 
three membered. Except for at the band intersections, the phase 
relationships cannot coincide among the three surfaces because of their 
topological differences.

The equivalence of partial Hamiltonians $h_{\bf k}$ between two surfaces 
at a band intersection implies a one-to-one correspondence between the 
subspaces of states of the same energy level. If the two sides of the 
surfaces are distinct, the correspondence would be simple so that one 
could expect a one-to-one correspondence between the IR's between the 
surfaces. However, as we see in Fig.\ref{BandIntersections2}, the 
correspondence is somewhat obscure due to the reduction of the primitive 
unit cells for P- and D-surfaces when the two sides are indistinguishable. 
The unfolding of the BZ's for these two surfaces results in the splitting 
of a single band intersection in the 6D BZ into two different special 
points (see TABLE \ref{T1}). The resulting two special points can either 
be (i) equivalent, as in the case of two $N$-points for P-surface, or 
(ii) inequivalent, as in the case of $\Gamma$- and $H$-points for P-surface.
For the case (i), a subspace of states at a band intersection is divided 
into two subspaces associated with the two {\bf k} points in the same star, 
so that the dimensionality of the IR's is reduced by half from that of the 
original subspace. Meanwhile, in the case (ii), a subspace of states at 
a band intersection is mapped onto a single subspace associated with 
either of the different special points. This argument is based on that 
each subspace at a single {\bf k} point is in general associated with 
a single IR, and that the total dimensionality of the original subspaces 
at a band intersection is identical between the relevant two surfaces.

We expect that there is a certain connection between the IR's of different 
surfaces at their band intersections. It appears that an IR of P-surface 
corresponds uniquely to an IR of D-surface, while an IR of G-surface 
corresponds to two IR's of P-surface as well as D-surface. 
In Fig.\ref{BandIntersections2} one finds that the IR $\Gamma_9$ of 
G-surface corresponds to either $\Gamma_9$ or $H_4$ of P-surface as well as 
to either $\Gamma_9$ or $R_4$ of D-surface. Interestingly, the connection 
between the IR's of P-surface at $P$-point and those of G-surface at 
$H$-point can be uniquely determined by the dimensionality of the IR's as
\begin{eqnarray*}
P_1(1),P_2(1) &\leftrightarrow& H_2(2),\\
P_3(2) &\leftrightarrow& H_3+H_4(2+2),\\
P_4(3), P_5(3) &\leftrightarrow& H_1(6),
\end{eqnarray*}
where the dimensionality of each IR is denoted by parenthesis. 
(Recall that the IR's $H_3$ and $H_4$ for G-surface is degenerate due to 
the time-reversal symmetry.)
For rigorous proofs of the general relationships between the IR's of the 
three surfaces, more detailed investigations on the group theoretical 
aspects of the problem may be required. Still, our results can be a starting 
point for further studies.

The relationship between the symmetry properties of eigenstates and the 
occurrence of nodal lines are established in Sec.\ref{symmetry}. 
The existence of nodal lines in an eigenstate necessitate a sufficient 
amount of kinetic energy, so that IR's with a large number of nodal lines 
cannot appear in a lowest part of the energy spectrum. In TABLES \ref{T2} 
and \ref{T3}, one can find that the following IR's accompany a significant 
number of nodal lines: \{$\Gamma_2$, $\Gamma_6$, $\Gamma_7$, $H_6$\} of 
P-surface and \{$\Gamma_2$, $R_1$, $R_2$, $R_6$\} of D-surface. 
These IR's are naturally absent in Figs.\ref{Pband_new} and \ref{Dband_new}. 
It is also seen that IR's with nodal lines appear roughly in the 
increasing order in the number of nodal lines as one goes up the energy 
axis. Therefore, nodal lines can be regarded as an important feature of 
eigenstates present in real-space in order to understand the overall 
characteristics of the systems.

For a full account of the electronic structures of the present systems, 
a more detailed study on the topological aspects of the problem should be 
important. Due to the topological characters of the surfaces, there may 
be a ``topological invariant'' associated with the continuum of eigenstates 
along a dispersion curve in the band structures. Such a topological 
invariant might be given as the phase winding numbers associated with 
characteristic closed paths on the surface. The topological characters of the 
eigenstates may explain the connectivity aspects of the energy bands. 
Further properties of the energy bands (e.g., the flatness, 
particle-/hole-type dispersions) might also be explained based on real-space 
pictures of the relevant eigenstates.

It is instructive to note that the arguments throughout this paper are not 
binded by the particular form of the curvature potential as long as it is 
determined intrinsically. Hence, by generalizing the potential to any 
function of the Gaussian curvature one may be able to deal with secondary 
effects of curvatures, such as the effect of varying thickness of the 
surface depending on the curvatures.

In summary, we have studied in detail the band structures of the strongly 
constrained electron systems onto P-, D-, and G-surfaces. 
The crystallographic relations of these surfaces in six dimensions have
proved to be useful in understanding the interrelations between the 
electronic structures of these surfaces. By introducing the hypothetical 
6D band structures, the coincidence of eigenstates among different surfaces at 
particular sets of {\bf k} points has been explained in terms of the idea of 
band intersections. We have given the exhaustive list of band intersections 
for the three surfaces. A numerical investigation of the band structures 
with a detailed analysis of the symmetry properties has also been performed. 
We have assigned IR's of the space groups to the energy bands displayed in 
Figs.\ref{Pband_new},\ref{Dband_new}, and \ref{Gband_new}. As a key feature
to understand the consequences of the symmetry of the eigenstates, the 
occurrence of characteristic nodal lines have been discussed somewhat in 
detail. We have given the conditions for nodal lines to be predicted by 
symmetry and also the exhaustive lists of IR's which entail nodal lines 
for P- and D-surfaces.

\begin{acknowledgments}
N. F. is greatly indebted to K. Niizeki and S. Takagi for helpful 
discussions and comments. He is also grateful to H. Aoki for useful 
communication. Financial supports from Japan Science and Technology 
Agency (JST) and Swedish Research Council (VR) are greatly acknowledged.
\end{acknowledgments}

\appendix*

\section{The index of $\Lambda^{\ast}_X\oplus\Lambda^{\ast}_Y$ in $\Lambda^{\ast}$}
\label{Appendix:Indices}

We consider the sublattices of $\Lambda^{\ast}$ given in the form 
$\Lambda_X^{\ast}\oplus\Lambda_Y^{\ast}$ ($X,Y=P,D$, or $G$). 
From Eq.(\ref{generatorLast}), it is straightforward to obtain the following:
\begin{eqnarray}
\Lambda^{\ast}_P\oplus\Lambda^{\ast}_D &=&
\left\{ ~i~(\vec{Q}_1+\vec{Q}_5-\vec{Q}_6) 
+ j~(\vec{Q}_2-\vec{Q}_4+\vec{Q}_6) 
+ k~(\vec{Q}_3+\vec{Q}_4-\vec{Q}_5)\right.\nonumber\\
 && \left. + ~l~(-\vec{Q}_2+\vec{Q}_3+\vec{Q}_4) 
+ m~(\vec{Q}_1-\vec{Q}_3+\vec{Q}_5) 
+ n~(-\vec{Q}_1+\vec{Q}_2+\vec{Q}_6)~\right\},\label{LPLD}\\
\Lambda^{\ast}_P\oplus\Lambda^{\ast}_G&=&
\left\{ ~i~(\vec{Q}_1+\vec{Q}_5-\vec{Q}_6) 
+ j~(\vec{Q}_2-\vec{Q}_4+\vec{Q}_6) 
+ k~(\vec{Q}_3+\vec{Q}_4-\vec{Q}_5)\right.\nonumber\\
 && \left. + ~l~(\vec{Q}_1-\vec{Q}_4-\vec{Q}_6) 
+ m~(\vec{Q}_2-\vec{Q}_4-\vec{Q}_5) 
+ n~(\vec{Q}_3-\vec{Q}_5-\vec{Q}_6)~\right\},\label{LPLG}\\
\Lambda^{\ast}_D\oplus\Lambda^{\ast}_G&=&
\left\{ ~i~(-\vec{Q}_2+\vec{Q}_3+\vec{Q}_4) 
+ j~(\vec{Q}_1-\vec{Q}_3+\vec{Q}_5) 
+ k~(-\vec{Q}_1+\vec{Q}_2+\vec{Q}_6)\right.\nonumber\\
 && \left. + ~l~(\vec{Q}_1-\vec{Q}_4-\vec{Q}_6) 
+ m~(\vec{Q}_2-\vec{Q}_4-\vec{Q}_5) 
+ n~(\vec{Q}_3-\vec{Q}_5-\vec{Q}_6)~\right\},\label{LDLG}
\end{eqnarray}
where $i$, $j$, $k$, $l$, $m$, and $n$ are arbitrary integers. 
In the following, the orthonormal basis sets of the 3D subspaces 
$E_P$, $E_D$, and $E_G$ are denoted as $\vec{e}_{P,i}$, $\vec{e}_{D,i}$ 
and $\vec{e}_{G,i}$ ($i=1,2,3$), respectively. These basis vectors are 
obtained as the row vectors of the corresponding orthogonal projectors, 
$\Pi_P$, $\Pi_D$, and $\Pi_G$.

\subsection{$\Lambda^{\ast}_P\oplus\Lambda^{\ast}_D$}

$\Lambda^{\ast}_P\oplus\Lambda^{\ast}_D$ is the sublattice of 
$\Lambda^{\ast}=\{\sum^6_{i=1}n_i\vec{Q}_i\}$ for which the coefficients 
$n_i$ are given by
\begin{eqnarray}
\left(\begin{array}{c}
n_1 \\ n_2 \\ n_3 \\ n_4 \\ n_5 \\ n_6
\end{array}\right)={\bf M}_{PD}
\left(\begin{array}{c}
i \\ j \\ k \\ l \\ m \\ n
\end{array}\right),\quad
{\bf M}_{PD}=\left(\begin{array}{cccccc}
1 & 0 & 0 & 0 & 1 & -1 \\
0 & 1 & 0 & -1 & 0 & 1 \\
0 & 0 & 1 & 1 & -1 & 0 \\
0 & -1 & 1 & 1 & 0 & 0 \\
1 & 0 & -1 & 0 & 1 & 0 \\
-1 & 1 & 0 & 0 & 0 & 1
\end{array}\right).
\end{eqnarray}
Since the index $|\det{\bf M}_{PD}|=4$, $\Lambda^{\ast}$ is divided into 
four sublattices equivalent to $\Lambda^{\ast}_P\oplus\Lambda^{\ast}_D$. 
Each sublattice is specified by a pair of indices, ($p,q$):
\begin{eqnarray}
p&=&n_1+n_3+n_4+n_6 \mod{2},\\
q&=&n_2+n_3+n_5+n_6 \mod{2}.
\end{eqnarray}
In particular, the sublattice $\Lambda^{\ast}_P\oplus\Lambda^{\ast}_D$ 
is specified by ($p=0$, $q=0$). It gives rise to the band intersection 
at $\Gamma$-point, $\vec{k}_P=\vec{k}_D=\vec{0}$, 
in Eq.(\ref{bandintersection}). The other three sublattices are obtained 
by translating $\Lambda^{\ast}_P\oplus\Lambda^{\ast}_D$ by $\vec{Q}_1$ 
($p=1,q=0$), $\vec{Q}_2$ ($p=0,q=1$), and $\vec{Q}_3$ ($p=1,q=1$). 
For each of these sublattices, the vectors $\vec{k}_P$ and $\vec{k}_D$ 
in Eq.(\ref{bandintersection}) can be taken as
\begin{eqnarray}
&&\vec{k}_P=(\vec{e}_{P,2}+\vec{e}_{P,3})/(2r\ell), \, 
\vec{k}_D=(\vec{e}_{D,2}-\vec{e}_{D,3})/(2s\ell)\nonumber\\
&&\qquad{\rm for}~ p=1, q=0 \quad (\vec{k}_P-\vec{k}_D=\vec{Q}_1),\\
&&\vec{k}_P=(\vec{e}_{P,1}+\vec{e}_{P,3})/(2r\ell), \, 
\vec{k}_D=(-\vec{e}_{D,1}+\vec{e}_{D,3})/(2s\ell)\nonumber\\
&&\qquad{\rm for}~ p=0, q=1 \quad(\vec{k}_P-\vec{k}_D=\vec{Q}_2),\\
&&\vec{k}_P=(\vec{e}_{P,1}+\vec{e}_{P,2})/(2r\ell), \, 
\vec{k}_D=(\vec{e}_{D,1}-\vec{e}_{D,2})/(2s\ell)\nonumber\\
&&\qquad{\rm for}~ p=1, q=1 \quad (\vec{k}_P-\vec{k}_D=\vec{Q}_3).
\end{eqnarray}
It is readily shown that these three band intersections are observed at the 
three membered star of M-point in the BZ of P-surface as well as at the 
three membered star of X-point in the BZ of D-surface, where the dodecagonal 
patches are taken as the primitive cells.

\subsection{$\Lambda^{\ast}_P\oplus\Lambda^{\ast}_G$}

A similar argument holds for $\Lambda^{\ast}_P\oplus\Lambda^{\ast}_G$,
for which
\begin{eqnarray}
{\bf M}_{PG}=\left(\begin{array}{cccccc}
1 & 0 & 0 & 1 & 0 & 0 \\
0 & 1 & 0 & 0 & 1 & 0 \\
0 & 0 & 1 & 0 & 0 & 1 \\
0 & -1 & 1 & -1 & -1 & 0 \\
1 & 0 & -1 & 0 & -1 & -1 \\
-1 & 1 & 0 & -1 & 0 & -1
\end{array}\right).
\end{eqnarray}
The index is $|\det{\bf M}_{PG}|=2$. Hence, there are two equivalent 
sublattices specified by an index, $r$:
\begin{eqnarray}
r=n_4+n_5+n_6 \mod{2}.
\end{eqnarray}
Apart from the trivial band intersection for $r=0$ at $\Gamma$-point, 
we have another band intersection for the sublattice $r=1$, which is 
obtained by translating $\Lambda^{\ast}_P\oplus\Lambda^{\ast}_G$ by 
$\vec{Q}_4$. The relevant vectors $\vec{k}_P$ and $\vec{k}_G$ can be 
taken as
\begin{eqnarray}
&&\vec{k}_P=(\vec{e}_{P,1}-\vec{e}_{P,2}+\vec{e}_{P,3})/(2r\ell), \, 
\vec{k}_G=\vec{e}_{G,1}/(2b\ell)\nonumber\\
&&\qquad{\rm for}~ r=1 \quad (\vec{k}_P-\vec{k}_G=\vec{Q}_4).
\end{eqnarray}
The band intersection is therefore observed at R-point in the BZ of 
P-surface as well as at H-point in that of the G-surface, where the 
dodecagonal patches are taken as the primitive cells.

\subsection{$\Lambda^{\ast}_D\oplus\Lambda^{\ast}_G$}

For $\Lambda^{\ast}_D\oplus\Lambda^{\ast}_G$,
\begin{eqnarray}
{\bf M}_{DG}=\left(\begin{array}{cccccc}
0 & 1 & -1 & 1 & 0 & 0 \\
-1 & 0 & 1 & 0 & 1 & 0 \\
1 & -1 & 0 & 0 & 0 & 1 \\
1 & 0 & 0 & -1 & -1 & 0 \\
0 & 1 & 0 & 0 & -1 & -1 \\
0 & 0 & 1 & -1 & 0 & -1
\end{array}\right).
\end{eqnarray}
The index is $|\det{\bf M}_{DG}|=1$. Therefore 
$\Lambda^{\ast}_D\oplus\Lambda^{\ast}_G=\Lambda^{\ast}$, and there is 
no extra band intersection other than the trivial one at $\Gamma$-point.




\newpage

\begin{thebibliography}{50}

\bibitem{SN91}
H. G. von Schnering and R. Nesper, Z. Phys. B: Condensed Matter {\bf 83}, 
407 (1991).

\bibitem{AHLL88}
S. Andersson, S. T. Hyde, K. Larsson, and S. Lidin, Chem. Rev. {\bf 88}, 
221 (1988).

\bibitem{JK71}
H. Jensen and H. Koppe, Ann. Phys. {\bf 63}, 586 (1971).

\bibitem{C81}
R. C. T. da Costa, Phys. Rev. A {\bf 23}, 1982 (1981).

\bibitem{C82}
R. C. T. da Costa, Phys. Rev. A {\bf 25}, 2893 (1982).

\bibitem{IN91}
M. Ikegami and Y. Nagaoka, Prog. Theor. Phys. Suppl. {\bf 106}, 235 (1991).

\bibitem{DE95}
P. Duclos and P. Exner, Rev. Math. Phys. {\bf 7}, 73 (1995).

\bibitem{CEK04}
G. Carron, P. Exner, and D. Krejcirik, J. Math. Phys. {\bf 45}, 774 (2004).

\bibitem{BV86}
 N. L. Balazs and A. Voros, Phys. Rep. {\bf 143}, 109 (1986).

\bibitem{AKTMK01}
H. Aoki, M. Koshino, D. Takeda, H. Morise, and K. Kuroki, 
Phys. Rev. B {\bf 65}, 035102 (2001).

\bibitem{KA05}
M. Koshino and H. Aoki, Phys. Rev. B {\bf 71}, 073405 (2005).

\bibitem{S75}
 M. Spivak, {\em A Comprehensive Introduction to Differential Geometry} 
(Publish or Perish, Berkeley, 1975), Vols. 3 and 4.

\bibitem{O86}
R. Osserman, {\em A Survey of Minimal Surfaces} (Dover, 1986).

\bibitem{BonnetFamily}
Different minimal surfaces which just differ by the Bonnet angles, 
$\alpha$, in Eq.(\ref{WEformula}) constitute a Bonnet family.

\bibitem{CS87}
 J. Charvolin and J.-F. Sadoc, J. Phys. (Paris) {\bf 48}, 1559 (1987).

\bibitem{SC89}
 J.-F. Sadoc and J. Charvolin, Acta Cryst. A {\bf 45}, 10-20 (1989).

\bibitem{Schwarz}
 H. Schwarz, {\em Gesammelte Mathematische Abhandlungen} 
(Springer, Berlin, 1890), vol.1.

\bibitem{Schoen}
 A. H. Sch$\ddot{\rm o}$n, {\em NASA Technical Report TN D-5541} 
(Washington DC, 1970).

\bibitem{RiemannSurface}
The Riemann surface consists of two sheets of complex plane with four 
branch cuts, and it corresponds to the unit cell of each TPMS.

\bibitem{GCMK00}
 P. J. F. Gandy, D. Cvijovi$\acute{\rm c}$, A. L. Mackay, and J. Klinowski, 
Chem. Phys. Lett. {\bf 314}, 543 (1999); 
P. J. F. Gandy and J. Klinowski, Chem. Phys. Lett. {\bf 321}, 363 (2000); 
Chem. Phys. Lett. {\bf 322}, 579 (2000).

\bibitem{(pqr)Triangle}
A ($p,q,r$) triangle is a triangle bounded by geodesics with the internal
angles, ($\pi/p,\pi/q,\pi/r$). Since the surfaces have negative Gaussian 
curvatures, the Gauss-Bonnet theorem requires the sum of these angles to 
be smaller than $\pi$.

\bibitem{OS93}
C. Oguey and J.-F. Sadoc, J. Phys. I France {\bf 3}, 839 (1993).

\bibitem{vsclassical}
In general, the curvature potential at any given point depends not only on the 
intrinsic curvatures but also on the extrinsic ones. Hence, the problem is 
essentially distinct from the classical counterpart, as discussed in 
Ref. \onlinecite{C81}).

\bibitem{SublatticeIndex}
This means that $\Lambda^{\ast}$ is divided into $\mu$ equivalent 
sublattices congruent to $\Lambda^{\ast}_X\oplus\Lambda^{\ast}_Y$.

\bibitem{Lax}
M. Lax, {\em Symmetry Principles in Solid State and Molecular Physics} 
(Wiley, New York, 1974).

\bibitem{TSPACE}
A. Yanase, {\em Fortran Program for Space Group (TSPACE)} 
(Shokabo, Tokyo, 1995).

\bibitem{ZakTables}
J. Zak, A. Casher, M. Gl$\ddot{\rm u}$ck, and Y. Gur, 
{\em The Irreducible Representations of Space Groups} 
(Benjamin, New York, 1969).

\bibitem{Herring}
C. Herring, Phys. Rev. {\bf 52}, 361 (1937).










\end{thebibliography}
\end{document}